\documentclass[conference]{IEEEtran}
\IEEEoverridecommandlockouts
\usepackage{cite}
\usepackage{amsmath,amssymb,amsfonts}
\usepackage{algorithmic}
\usepackage{graphicx}
\usepackage{textcomp}
\usepackage{xcolor}
\usepackage{booktabs}
\usepackage{adjustbox}
\usepackage{hyperref}
\def\BibTeX{{\rm B\kern-.05em{\sc i\kern-.025em b}\kern-.08em
    T\kern-.1667em\lower.7ex\hbox{E}\kern-.125emX}}

\definecolor{extra-light-gray}{rgb}{0.95,0.95,0.95}
\usepackage{mdframed}
\usepackage{framed}
\usepackage{environ}
\NewEnviron{opus-box}[1][]{%
\vspace{0.1cm}
\begin{mdframed}[backgroundcolor=extra-light-gray]
    \BODY
\end{mdframed}
\vspace{0.1cm}
}

\begin{document}

\title{Insights on Microservice Architecture Through the Eyes
of Industry Practitioners%\\\thanks{Conselho Nacional de Desenvolvimento Científico e Tecnológico - CNPq}
}

\author{%
\IEEEauthorblockN{Vinicius L. Nogueira}
\IEEEauthorblockN{Fernando S. Felizardo}
\IEEEauthorblockN{Aline M. M. M. Amaral}
\IEEEauthorblockA{\textit{Department of Informatics} \\
\textit{Maringá State University}\\
Maringá, Brazil} %\\
\and

\IEEEauthorblockN{Wesley K. G. Assunção}
\IEEEauthorblockA{\textit{Department of Computer Science} \\
\textit{North Carolina State University}\\
Raleigh, United States of America} % \\
%wguezas@ncsu.edu}

\and

\IEEEauthorblockN{Thelma E. Colanzi}
\IEEEauthorblockA{\textit{Department of Informatics} \\
\textit{Maringá State University}\\
Maringá, Brazil} %\\
}

\maketitle

\begin{abstract}
The adoption of microservice architecture has seen a considerable upswing in recent years, mainly driven by the need to modernize legacy systems and address their limitations.
Legacy systems, typically designed as monolithic applications, often struggle with maintenance, scalability, and deployment inefficiencies. This study investigates the motivations, activities, and challenges associated with migrating from monolithic legacy systems to microservices, aiming to shed light on common practices and challenges from a practitioner's point of view. 
We conducted a comprehensive study with 53 software practitioners who use microservices, expanding upon previous research by incorporating diverse international perspectives. Our mixed-methods approach includes quantitative and qualitative analyses, focusing on four main aspects: (i) the driving forces behind migration, (ii) the activities to conduct the migration, (iii) strategies for managing data consistency, and (iv) the prevalent challenges.
Thus, our results reveal diverse practices and challenges practitioners face when migrating to microservices. Companies are interested in technical benefits, enhancing maintenance, scalability, and deployment processes. Testing in microservice environments remains complex, %with a focus on unit and integration testing, 
and extensive monitoring is crucial to managing the dynamic nature of microservices. Database management remains challenging. %, with companies divided between centralized and decentralized approaches. 
While most participants prefer decentralized databases for autonomy and scalability, challenges persist in ensuring data consistency. Additionally, many companies leverage modern cloud technologies to mitigate network overhead, showcasing the importance of cloud infrastructure in facilitating efficient microservice communication.
\end{abstract}


% ------------------------------------------------------
% ------------------------------------------------------
% ------------------------------------------------------

\section{Introduction}

% \textbf{Context:} \wk{What is this paper about}

% \textbf{Problem:} \wk{What are the practical challenges}

% \textbf{Related work / Limitations:} \wk{What others have done, but what still needs to be done.}

% \textbf{Goal / Methodology:} \wk{What we do, and how do we address the limitation above?}

% \textbf{Summary of the results:} 

% \textbf{Contributions:} \wk{how our results help developers/companies/researchers?}

Legacy software systems often present challenges in meeting the demands of modern business~\cite{Seacord2003, abgaz2023decomposition}. Since monolithic architecture had been the norm for decades, most of the existing software systems adopt such an architecture~\cite{stojkov2021}. 
However, the very nature of large-size monoliths presents several challenges, in which their inherent complexity makes maintenance and evolution cumbersome~\cite{Dragoni2017}. For instance, the deployment of monolithic applications tends to be suboptimal, and they also have limited scalability and technological flexibility~\cite{Dragoni2017}.

To address these problems and take advantage of enticing promises made by the rise of cloud computing~\cite{jamshidi2013}, companies are turning to microservice architecture as a means to modernize their legacies~\cite{chen2017, taibi2017, abgaz2023decomposition, Carvalho2024}. Microservice architecture is an architectural style that emphasizes the splitting of an application into small and lightweight services, which are specifically built to perform a very cohesive business capability~\cite{Lewis2014}. This style is an evolution of traditional service-oriented architecture~\cite{erl2005serviceoriented}. Architectures such as microservices carry valuable characteristics that help with the limitations of monolithic systems, such as easier maintenance, improved scalability, and partial deployment~\cite{henry2020}, and allow users to capitalize on the advantages offered by cloud environments~\cite{balalaie2016-2}.

Despite the benefits of a service-oriented architecture, modernizing a monolith through microservices remains challenging\cite{abgaz2023decomposition, Wolfart2021}. There is no one-size-fits-all architecture that can be applied universally, and therefore, each microservice architecture must be tailored to meet the specific needs of an application~\cite{henry2020, Wolfart2021}. Several studies explored the migration of legacy systems to microservice architecture \cite{bucchiarone2018, cruz2019, hayretci2021, Wolfart2021, abgaz2023decomposition}. However, there is still a limited amount of in-depth research on the practical aspects for conducting the modernization.

To gain a practical view of modernization with microservices, Zhou et al.~\cite{zhou2023} conducted interviews with practitioners from 20 companies to investigate the practical implications of microservices architecture. Their findings show the discrepancies between idealized visions and real-world practices concerning this architectural style, and emphasize the importance of informed decision-making and the need to proactively address potential challenges in adopting microservices. 
Other papers also present surveys with practitioners~\cite{taibi2017, fritzsch2019,difrancesco2018}. However, while these papers explore strategies and challenges for migration, they do not provide in-depth insights into how practitioners deal with certain issues. These studies often focus on a single approach, some using quantitative data \cite{difrancesco2018}, and others are qualitative studies \cite{taibi2017,fritzsch2018}. 

Even though existing studies indicate organizational challenges, including mindset shifts towards agile processes and team collaboration, they do not fully address the entire modernization process. For instance, little is known about %including the complexities of 
data model transformation and the integration of microservices within existing infrastructure. These gaps highlight the need for further research combining qualitative and quantitative analyses to provide a more comprehensive understanding of microservices migration in practice.
In a previous study, we conducted a preliminary survey with 56 practitioners, primarily from Brazil, of whom 35 use microservices~\cite{Colanzi2021Original}. However, we predominantly evaluated the technical aspects of the modernization process and investigated (in a \textit{confirmatory} study) whether industry practices were aligned with academic principles based on the roadmap presented by Wolfart et al.~\cite{Wolfart2021}, ultimately confirming this alignment.

Motivated by the above-mentioned limitations, we investigate a different perspective from our previous work. In this study, we adopt a more \textit{exploratory} approach to uncover how professionals navigate the complexities of microservice migration in their organizations. In addition to the common technical, organizational, and operational perspectives of microservice migration, we take a more practical approach to investigate how and why practitioners deal with challenges related to data consistency and common pitfalls to avoid during migration.
Other limitations of existing studies is that they usually have a few dozens (around 20) of participants~\cite{taibi2017, fritzsch2019,difrancesco2018} or are geographically limited to a single region~\cite{fritzsch2019}. To diversify our sample and obtain more heterogeneous results, our survey includes more participants from different regions. We increased the total number of microservice users from 35 to 53, with 18 new responses. Using this newly collected data, we explore the motivations, activities, and challenges of migrating from monolithic legacy systems to microservices.

More specifically, our study consists of quantitative and qualitative analyses of data gathered from a survey-based study involving software practitioners. We essentially want to better understand \textit{why} and \textit{how} practitioners make certain decisions when modernizing their systems. We investigate four key aspects of the legacy system modernization process: (i) the driving forces compelling companies to migrate, (ii) the methodologies employed by these companies in their modernization process, (iii) the strategies implemented to handle database and data consistency concerns, and (iv) the prevalent challenges encountered throughout the modernization.

The results reveal diverse motivations for migration, such as the need for improved scalability and maintainability, confirming previous findings \cite{fritzsch2019, taibi2017, wang2020}. We identified a range of methodologies and strategies practitioners employ, particularly in managing data consistency and ensuring efficient deployment. To ensure data consistency, sophisticated techniques such as Command Query Responsibility Segregation (CQRS) and Event Sourcing were mentioned when working with decentralized systems. Additionally, efficient deployment is often achieved through robust monitoring tools, automatic monitoring mechanisms, and leveraging modern cloud technologies to mitigate network overhead and guarantee smooth operation across microservices. Furthermore, we discuss some challenges, including the complexity of testing microservice environments and the ongoing debate between centralized and decentralized database management approaches. The findings reaffirm the critical role of cloud infrastructure in facilitating effective microservice communication and indicate areas where further research and tailored solutions are needed.

The main contributions of our work include providing a comprehensive view of the current practices and challenges industry that practitioners face when migrating to microservices. The insights gained from this study can help developers and companies make informed and cost-efficient decisions when considering microservice adoption. By documenting real-world experiences and identifying common obstacles, our research offers valuable guidance for overcoming the hurdles associated with microservice implementation. Moreover, the study lays the groundwork for future research, particularly in areas such as testing methodologies and database management in microservice architectures, thereby supporting the continued evolution of best practices in this domain.

%\textcolor{blue}{In this work, our goal is to \textit{capture the perceptions of practitioners who work with microservices architecture daily, gathering their insights into the benefits and challenges of its adoption}. Their first-hand experiences allow us to understand better the practical implications of modernizing monolithic legacy systems with microservices. With these insights, we hope to help practitioners make more cost-efficient decisions when addressing challenges such as difficult maintenance and scalability. Our participants' insights are grounded in real-world applications and issues rather than theoretical assumptions, making their thoughts highly relevant and practical.}

The remainder of this paper is structured as follows. %Section \ref{sec:background} provides background information relevant to the study. 
Section~\ref{sec:related} reviews related work. %, situating our research within the existing literature. 
Section \ref{sec:methods} details the study design and execution, with our research methodology and data collection process. Section \ref{sec:results} presents the results and discussion. %Based on this section's findings, we report the discoveries that answer our research questions.
Section \ref{sec:lessons} contains extracted practical lessons for practitioners.
Section \ref{sec:threats} addresses the threats to the validity of our study and Section \ref{sec:conclusion} concludes the paper by summarizing the main findings and suggesting directions for future research.

% ------------------------------------------------------
%\section{Background}
%\label{sec:background}

{\color{blue}
}

% ------------------------------------------------------
% ------------------------------------------------------
% ------------------------------------------------------

\section{Related Work}
\label{sec:related}

Wolfart et al.~\cite{Wolfart2021} present a comprehensive roadmap for modernizing monolithic legacy systems using microservices. The proposed roadmap comprises eight activities grouped into four phases: initiation, planning, execution, and monitoring. To understand how monolithic legacy systems are migrated to microservices, the authors used a systematic mapping method based on 62 studies to extract information regarding the motivations to start a modernization, the activities used to conduct the modernization, and the information used as input and generated output for each of those activities. The main contribution of their study includes providing a roadmap for practitioners to plan, execute, and monitor the modernization process, serving as a reference for researchers to design new studies, and motivating tool builders to address existing needs.

Using Wolfart et al.'s roadmap, in a previous study, we investigated whether industry and academia are aligned in their approaches to modernizing legacy systems using microservices~\cite{Colanzi2021Original}. We surveyed 35 software practitioners who use microservices, and the results indicate that, for the most part, companies follow best practices shown in the literature, performing activities outlined in the modernization roadmap~\cite{Wolfart2021}. However, the responses show that data persistence is still controversial among practitioners, with many companies using decentralized databases and managing data consistency through application logic. Additionally, in our prior work, the analyses were focused on technical aspects, such as the integration and communication tools the participants used.
Overall, that study concludes that industry and academia are generally well-aligned. However, some areas in the context of microservices still needed further research. For instance, our survey did not include questions regarding the process of understanding the legacy system before migration, even though most participants said they performed this activity. 

More recently, Zhou et al.~\cite{zhou2023} conducted an empirical study to explore the divergences between the ideal concepts of microservices architecture and real industry practices. Through interviews with practitioners from 20 software companies, the study identified common practices and challenges in implementing microservices. They synthesized their findings into eight pairs of practices and pains and extracted five critical decision-making aspects to help balance the benefits and drawbacks of the architecture. The authors conclude that (i) the benefits of microservices are confirmed by practitioners; and (ii) the potential problems must be carefully addressed, including systematic evaluation, organizational transformation, decomposition, distributed monitoring, and bug localization, to optimize the implementation of microservice architectures.

Di Francesco et al.~\cite{difrancesco2018} surveyed 18 practitioners to gather quantitative information about migrations towards microservices architecture. Their results emphasize the importance of well-planned strategies and effective team collaboration. Among the challenges identified, the authors reinforce the migration of pre-existing data from legacy databases to microservices as a significant open problem. Sixty percent of their participants said they maintained their original database structures even after migrating. 
Similarly, Fritzsch et al.~\cite{fritzsch2019} conducted interviews with 16 professionals from 10 German-based companies, focusing on intentions, strategies, and challenges. Both of these studies reveal significant organizational challenges. However, these studies provide either qualitative or quantitative insights but leave open questions regarding the approaches practitioners use to tackle challenges. 

Our study addresses these gaps by incorporating qualitative and quantitative analyses from a larger and more diverse sample of 53 practitioners from various international backgrounds. We report on real-world experiences associated with migrating to microservices, covering the entire process from the initial motivations for migration through the migration itself to the strategies practitioners employ for dealing with challenges such as maintaining data consistency and ensuring efficient deployment. By doing so, our research provides a more holistic understanding of the practical realities of microservices adoption and offers actionable insights for professionals.

% Despite the evident benefits of migrating to microservices architecture, there remains a gap in research concerning best practices and strategies for achieving a smooth modernization, balancing maximizing advantages while minimizing the potential weaknesses inherent in this architecture. This paper aims to contribute insights from industry practitioners who have worked during migrations or on migrated systems and use their experiences to help architects avoid pitfalls and succeed when using this technology.

% \wk{point what are the limitations that need another study. What we do different from related work.}

% ------------------------------------------------------
% ------------------------------------------------------
% ------------------------------------------------------

\section{Study Design and Execution} 
\label{sec:methods}

This section presents a detailed description of our study.

% \wk{Study Goal}

% \wk{Reserach questions with rationale of the differences from previous work}

% \wk{Questionaire}

% \wk{Characterization of Participants}

% \wk{Data analysis}

\subsection{Goal and Research Questions}

% The main goal of this study is to provide information to professionals to allow them to make informed decisions when migrating to microservices. By analyzing the responses given by participants, we offer an overview of the current practices and challenges industry practitioners face. We translated our main research goal into the following four research questions. 

% The main goal of this study is to provide information to professionals to help them make informed decisions when migrating to microservices. To achieve this goal, we conducted a multi-method approach, performing both quantitative and qualitative analyses of our survey data.

The main goal of our study is to \textit{understand \textbf{why} and \textbf{how} industry practitioners make their decisions when migrating monolithic legacy systems}. This understanding can help other professionals make informed decisions when performing similar activities. To this end, we used a multi-method approach, performing quantitative and qualitative data analyses.

When talking about microservice migration, existing work usually focuses on technical \cite{Colanzi2021Original, zhou2023}, economic \cite{taibi2017} or even organizational \cite{fritzsch2019, zhou2023, difrancesco2018 } aspects of the migration process, but more importantly, these researches predominantly focus on identifying what goes into these concepts.
In contrast, we aim at a more practical perspective, leveraging the real-life experiences of practitioners to go further and understand how and why industry professionals do or do not make specific decisions when modernizing legacy systems with microservices.

We translated our main research goal into four research questions (RQ), as follows: 

\vspace{1mm} %\noindent
\textbf{RQ1. Why do companies migrate monolithic legacy systems to microservices?} By answering this research question, we wanted to understand why practitioners/companies migrated to microservice architectures and the motivations driving companies to adopt microservices. This can help practitioners assess and align the potential benefits with their organizational goals.

\vspace{1mm} %\noindent
\textbf{RQ2. How do practitioners approach the migration process? } We aimed to understand participants' strategies to execute the migration successfully. This question seeks to identify and characterize the activities employed during the migration process, establishing a basis for the migration.

\vspace{1mm} %\noindent
\textbf{RQ3. What are the concerns regarding data persistence in modernizing legacy systems with microservices? } We address some of the pressing issues regarding the recurrent topic of data management in migration. Addressing this question will detach critical issues related to this topic.%, offering insights into best practices for maintaining data consistency and integrity during and after the migration.

\vspace{1mm} %\noindent
\textbf{RQ4. What are the main challenges and pitfalls to avoid when migrating legacy systems to microservices? } We studied the common challenges and pitfalls faced by practitioners. By answering this question, we aim to equip professionals with knowledge to anticipate and mitigate similar problems, leading to a more successful migration.

\subsection{Design of the Questionnaire}

% In this study, \textcolor{red}{we analyze the answers to a survey initially proposed by Colanzi et al. \cite{Colanzi2021Original}; with this new analysis, we aim to observe the modernization processes from the perspective of employees of software development companies.} The original questionnaire had 56 responses, with 35 participants using microservice architecture. This paper expanded the survey to include participants from multiple countries worldwide, adding 27 new responses. Of these, 18 were from microservice users, increasing our original sample size by more than 50\%. Hence, we now have 53 responses from microservice users.

In this study, we performed survey-based research to observe the modernization processes from the perspective of software development companies' practitioners. For that, we designed a questionnaire composed of 29 questions, of which 20 are close-ended, and 9 are open-ended questions. The open-ended questions were included so that participants could explain their answers in detail. Due to space limitations, the questionnaire is available in our supplementary material~\cite{rep2024}. We used a Google Form\footnote{\url{https://www.google.com/forms/about/}} to collect practitioners' responses.
%The questionnaire is composed of 29 questions. 20 are close-ended, and 9 are open questions so that participants can explain their answers in detail. The questions and participants' responses are anonymously available at~\cite{rep2024}.

Based on the migration process, the questions are grouped into subjects of interest (or categories). These subjects are: (i) Driving forces for migration, question Q1; (ii) Execution of the migration process, questions Q2 to Q10; (iii) Database concerns, questions Q11 to Q21; and (iv) Challenges and difficulties, questions Q22 to Q29. These four categories are designed to help answer a corresponding research question. Additionally, the questionnaire has questions used to characterize the participants (i.e., demographic questions). 
%With the responses collected, we then grouped different questions into subjects of interest (or categories) in the migration process, and these subjects are: i) Driving forces for migration - \textbf{question}: Q1; ii) Execution of the migration process - \textbf{questions}: Q2 - Q10; iii) Database concerns - \textbf{questions:} Q11 - Q21; and iv) Challenges and difficulties - \textbf{questions:} Q22 - Q29. These four categories are designed to help answer a corresponding research question.

%We used a Google Form\footnote{\url{https://www.google.com/forms/about/}} to collect participants' responses. The questionnaire contained close-ended and open-ended questions about activities related to migrating legacy systems to microservice architecture, and questions used to characterize the participants. 

In our prior survey~\cite{Colanzi2021Original}, we noted that many companies perform the activity of ``Understanding the legacy system'' during modernization. However, since the original questionnaire did not have in-depth questions on this topic, two new questions were added regarding how companies understood their original system before modernization. These two questions only have 18 responses from the new participants, as explained below.

The questionnaire of our preliminary study~\cite{Colanzi2021Original} received 56 responses, with 35 participants using microservice architecture, which was considered in our study. This present work expanded the survey to include participants from multiple countries worldwide, adding 27 new responses. Of these, 18 were from microservice users, increasing our original sample size by more than 50\%. Hence, our present study has a set of 53 responses from microservice users, used as a source of information to answer the RQs.

% *****Tirei a tabela para reduzir espaço e coloquei as informações das questões no parágrafo acima****
%Table \ref{tabelaQ} shows which questions relate to the different categories.

%\begin{table}[ht]
%\centering
%\caption{Questions related to each category}
%\footnotesize
%\begin{tabular}{ll}
%\hline
%\textbf{Subject of Interest}       & \textbf{Related Questions} \\ \hline
%Driving Forces for Migration       & Q1 \\%[2mm]
%Execution of the Migration Process & \begin{tabular}[c]{@{}l@{}}Q2 - Q10\end{tabular} \\%[2mm]
%Database Concerns                  & \begin{tabular}[c]{@{}l@{}}Q11 - Q21\end{tabular} \\%[2mm]
%Challenges and Difficulties        & \begin{tabular}[c]{@{}l@{}}Q22 - Q29\end{tabular}  \\ 
%\hline
%\end{tabular}
%\label{tabelaQ}
%\end{table}

% An illustration of the used procedure is shown in Figure \ref{designchart}.

% \begin{figure}[ht]
%     \centering
%     \includegraphics[scale=0.35]{images/Flowchart.png}
%     \caption{Experimental procedure}
%     \label{designchart}
% \end{figure}

% \subsection{Data Analysis}

% ------------------------------------------------------
% ------------------------------------------------------
% ------------------------------------------------------

\subsection{Search and Characterization of Participants}
\label{sec:characterization}

To find participants, we first looked for researchers who published papers related to microservices architecture; however, this strategy resulted in low participation. As a second approach, we posted on different social media websites (LinkedIn, X---former Twitter, and Facebook) and contacted students and former colleagues from different countries. This strategy proved much more successful, reaching more practitioners and enabling us to collect enough answers to our questionnaire.

% As previously mentioned, this paper builds upon the research results by Colanzi et al. \cite{Colanzi2021Original}, which surveyed 35 microservice users. To diversify and expand our sample size, we collected an additional 18 responses.

Adding our previous data to the newly collected replies, we now have a sample of 53 responses. Among the geographic distribution of the participants: 41 (77.4\%) of those responses are from Brazil, and 12 (22.6\%) from different countries, spread as follows: Ireland 2 (3.8\%), England 2 (3.8\%), Spain 2 (3.8\%), Uruguay 2 (3.8\%), Argentina 1 (1.9\%), Australia 1 (1.9\%), Portugal 1 (1.2\%) and United States of America (USA) 1 (1.2\%) response.

% As this paper builds upon the results of the research by Colanzi et al. \cite{Colanzi2021Original}, which had 56 responses, 27 responses were obtained with the newly proposed survey.
% We now have a total sample of 83 responses, including the data. 64 (77.1\%) of those responses are from Brazil, and 19 (22.9\%) from different countries, spread as follows: Ireland 3 (3.6\%), England 2 (2.4\%), Spain 2 (2.4\%), Uruguay 2 (2.4\%), Argentina 1 (1.2\%), Australia 1 (1.2\%), Canada 1 (1.2\%), Germany 1 (1.2\%), Hungary 1 (1.2\%), Iran 1 (1.2\%), Italy 1 (1.2\%), Luxembourg 1 (1.2\%), Portugal 1 (1.2\%) and United States of America (USA) 1 (1.2\%) response.

Most participants are Software Engineers, namely 25 (47.2\%). Other positions include Software Architect 8 (15.1\%), Project Manager 7 (13.2\%), Software Analyst 5 (9.4\%), Researcher 2 (3.8\%), Software Tester 2 (3.8\%), Software Specialist 2 (3.8\%), CEO 1 (1.9\%), and Innovation Director 1 (1.9\%).
The majority (29 or 54.7\%) of the companies where participants work defines their hierarchical structure considering the business area. At the same time, 18 (34\%) use the development area, and the remaining 6 (11.3\%) use a mixed structure of business and development.

Our participants bring diverse experiences to this study. Regarding overall experience with software development, 24 (45\%) participants have between 4 to 10 years of experience, while 10 (19\%) have 11 to 15 years of experience. A smaller group of 7 (13\%) participants has 16 to 20 years of experience, and another 7 (13\%) have been in the field for over 20 years. Notably, 27 (51\%) of the participants have been with their current company for 3 years or less, 15 (28\%) have worked there for 4 to 8 years, and 2 (4\%) have been in the same company for over 15 years. Regarding experience with microservices architecture migration, 26 (49\%) of the participants reported having some experience (5 years or less), while 8 (15\%) are considered experienced users with 5 to 8 years of experience. However, 6 (11\%) participants have no experience in MSA migration. As these questions were not mandatory, some participants did not respond. Table \ref{tab:Experience} illustrates the total responses.

\begin{table}[!tp]
\centering
\caption{Experience Demographics}
\footnotesize
\begin{adjustbox}{width=1\linewidth}
\begin{tabular}{lr}

\hline
\textbf{Experience}                                   & \textbf{Number (\%)} \\ \hline
\textbf{Overall experience with software development} &                                      \\
4 to 10 years experience                              & 24 (45\%)                            \\
11 to 15 years experience                             & 10 (19\%)                            \\
16 to 20 years experience                             & 7 (13\%)                             \\
More than 20 years experience                         & 7 (13\%)                             \\
No response                                           & 5 (9\%)                             \\
\hline
\textbf{Time working with current company}            &                                      \\
3 years or less                                       & 27 (51\%)                            \\
4 to 8 years                                          & 15 (28\%)                            \\
9 to 15 years                                         & 4 (8\%)                              \\
More than 15 years                                    & 2 (4\%)                              \\
No response                                           & 5 (9\%)                             \\
\hline
\textbf{Experience with MSA migration}                &                                      \\
No experience                                         & 6 (11\%)                             \\
5 years or less of experience (Some experience)       & 26 (49\%)                            \\
5 to 8 years experience (Experienced user)              & 8 (15\%)                             \\
No response                                           & 13 (25\%)                            \\ \hline
\end{tabular}
\end{adjustbox}
\label{tab:Experience}
\end{table}

\subsection{Data Analysis}
\label{sec:data_analysis}

We analyzed the responses in each category to answer the proposed research questions. For the close-ended questions, we performed a quantitative analysis. For the open-ended ones, we performed a thematic synthesis following the recommendations presented by Cruzes \& Tore~\cite{cruzes2011}.

% ------------------------------------------------------
% ------------------------------------------------------
% ------------------------------------------------------

\section{Results and Discussion}
\label{sec:results}

%\textcolor{blue}{This section presents the results of our analysis considering the 53 participants.} 
The results are presented in four subsections, one for each category, and these subsections show our extracted quantitative and qualitative data and answer their respective research question.

%Of our 83 participants, 30 said their company does not work with microservice architecture. Since this research aims to evaluate the usage aspects of microservice architecture in software development companies, the results presented here refer only to the 53 (18 for our two new questions) companies that use it.

% -- -- -- -- -- -- -- -- -- -- -- -- -- -- -- -- --

\subsection{RQ1 - Driving Forces for Migration}
\label{1.Drivingforces}

According to the literature~\cite{Wolfart2021}, the modernization of legacy systems with microservices is driven by various motivating factors. However, to understand the driving forces that lead companies to migrate their systems in practice, we asked participants (a multiple-choice question) \textit{why} they decided to modernize legacy systems through microservice architectures.

%\subsubsection{Quantitative Analysis}
Table~\ref{tab:Driving_Forces} presents the selected driving forces ranked by popularity. The results indicate a slight preference towards technical and operational motivations, although participants considered all perspectives. The responses reflect a balanced interest in the benefits microservices offer.
Specifically, the top three driving forces identified are consistent with those mentioned in existing literature \cite{fritzsch2019, Wolfart2021}: \textit{Easier maintenance and evolution} (81.1\%), \textit{Optimized scalability} (79.2\%), and \textit{Independent and automated deploy} (75.4\%). These driving forces were cited by more than 70\% of participants, indicating their importance in the decision to migrate.

\begin{table}[!tp]
\centering
\caption{Driving Forces}
\footnotesize
\begin{adjustbox}{width=1\linewidth}
\begin{tabular}{l|l|l}

\hline
\textbf{Driving Force}                    & \textbf{Total} (\%)   & \textbf{Perspective}    \\ \hline
Easier maintenance and evolution          & 43 (81.1\%)  & Technical      \\
Optimized scalability                     & 42 (79.2\%)  & Operational    \\
Independent and automated deploy          & 40 (75.4\%)  & Operational    \\
Technology flexibility                    & 29 (54.7\%)  & Technical      \\
Team Autonomy                             & 26 (49.1\%)  & Operational \\
Loosely coupled services                  & 25 (47.1\%)  & Technical      \\
Reduced time to market                    & 25 (47.1\%)  & Organizational \\
Agility enabler                           & 18 (33.96\%) & Organizational \\
Cohesive services                         & 17 (32.1\%)  & Technical      \\
Infrastructure facilities                 & 14 (26.4\%)  & Operational    \\
Easier reuse                              & 12 (22.7\%)  & Technical      \\
Domain-driven design delimited contexts   & 1 (1.9\%)    & Technical      \\
Better utilization of cloud computing resources & 1 (1.9\%)    & Technical      \\
Better operational autonomy                           & 1 (1.9\%)    & Operational \\ \hline
\end{tabular}
\end{adjustbox}
\label{tab:Driving_Forces}
\end{table}

Other significant motivations include \textit{Technology flexibility} (54.7\%) and \textit{Reduced time to market} (47.1\%). These factors suggest that the ability to adapt and innovate rapidly is a common consideration for organizations. Additionally, \textit{Team Autonomy} (49.1\%) and \textit{Loosely coupled services} (47.1\%) point towards a desire for greater independence and modularity within development teams.

Less commonly cited driving forces include \textit{Agility enabler} (33.9\%), \textit{Cohesive services} (32.1\%), \textit{Infrastructure facilities} (26.4\%), and \textit{Simpler Reuse} (22.6\%). Also, few participants used the ``others'' option to suggest their particular driving forces, naming unique reasons such as \textit{better operational autonomy}, \textit{better utilization of cloud computing resources}, and \textit{domain-driven design delimited contexts} (cited by only 1 participant each, 1.9\%).

%\wk{o mesmo participante respondeu essas categorias, ou foram 1 de participantes diferentes?}.

\begin{opus-box}
\textbf{Answering RQ1:} 
The primary driving forces for adopting microservices are predominantly operational and technical, focusing on enhancing maintenance, scalability, and deployment processes. This is in accord with the broader industry trends~\cite{zhou2023, taibi2017, bucchiarone2018}, which emphasize the technical and operational advantages of microservices in software development, such as easier maintenance/evolution and the possibility of using multiple technologies~\cite{carvalho2019, Chen2018}.
%However, given that industry and academia are in alignment in most aspects of migration \cite{Colanzi2021Original}, the prevalence of technical perspectives motivating migration can be attributed to the strong emphasis in the existing literature on the technical benefits of microservices~\cite{carvalho2019, Chen2018}. As a result, professionals tend to prioritize these technical aspects, as there is a lesser focus on the organizational aspects of modernization in currently available research \cite{Wolfart2021}.

\end{opus-box}

% In summary, the analysis reveals that the primary driving forces for adopting microservices are predominantly operational and technical, focusing on enhancing maintenance, scalability, and deployment processes. This aligns with the broader industry trends \cite{zhou2023, taibi2017, bucchiarone2018}, emphasizing the technical and operational advantages of microservices in software development, such as easier maintenance/evolution and the possibility of using multiple technologies \cite{carvalho2019, Chen2018}.

% However, the prevalence of technical perspectives might be because many studies advocate mostly technical benefits of microservices\cite{carvalho2019, Chen2018}. And existing work tend to focus less on the organizational aspects of modernization \cite{Wolfart2021}.

%\subsubsection{Answering RQ1}

% -- -- -- -- -- -- -- -- -- -- -- -- -- -- -- -- --

\subsection{RQ2 - Performing the Modernization}

\label{2.understanding}

%To effectively migrate legacy systems to microservices, understanding the existing architecture is an important first step \cite{Wolfart2021}. 

To understand \textit{how} companies carry out the migration process, we asked five multiple-choice and three open-ended questions. First, participants were asked about their preferred activities or strategies to understand their legacy systems before migration. This is one of the newly introduced questions.

%\subsection{Quantitative Analysis}
The most common strategy chosen was \textit{A study with the most experienced people that worked in the development of the legacy system}, selected by 10 out of 18 (55.6\%) participants. This result aligns with the findings by Taibi et al.\cite{taibi2017}, suggesting that involving experienced developers can significantly reduce the effort needed to develop microservice systems.
\textit{Domain-Driven Analysis} (50\%) and \textit{Reading system documentation} (44\%) were also frequently cited. These activities are well-regarded in the literature for their effectiveness in software development. Domain-driven design is a very flexible methodology \cite{fowler2020}, crucial for defining the boundaries and responsibilities of each created microservice \cite{velepucha2023}, while thorough documentation provides additional information in any project~\cite{aghajani2020}.
\textit{Dynamic Analysis} (39\%) is a fairly mentioned strategy, reflecting a need to understand runtime behavior, which can be a concern during modernization~\cite{junior2019}.

We also asked an open-ended question regarding the activities performed to understand the legacy system, and using thematic analysis to analyze our 18 responses, we arrived at four major themes for this subject.

The first theme, \textbf{Understanding Through Decoupling and Monitoring}, illustrates that participants perform the modernization of legacy systems through strategies such as decoupling components and thorough monitoring. Participant P67 reinforced the importance of breaking down the system into manageable parts and connecting them to new microservices while maintaining the legacy system until stability is ensured: ``\textit{As far as I'm aware, the modernization occurs by understanding the legacy system into components that could be decoupled [...] we execute the modernization, and slowly connect consumers to the new microservices while still having the legacy. Once they are stable, the legacy system can be decommissioned.}''

As a second theme, \textbf{Importance of Documentation} also plays an important role, as emphasized by P68: ``\textit{The development team had several meetings with the legacy application team, in which the code was presented, and many UML diagrams were specified to document existing legacy application and support the understanding.}'' The emphasis on documentation ensures that there is a clear and shared understanding of the system’s structure and behavior.

Two other themes are \textbf{Domain-Driven Design (DDD)} and \textbf{Effective Collaboration and Communication}, both mentioned by participants as key concepts in understanding the legacy system. Participants P69, P75, and P82 pointed out the value of DDD in comprehending business complexities and planning migration. The response of P69 addresses both of those themes: ``\textit{DDD techniques were very useful to understand the business complexities and core domain. To plan and execute the migration, activities mainly involved keeping close contact with the development/maintenance teams [...].}''
Team meetings and advice given by more experienced practitioners were also cited by participants P73, P76, and P78, who said: ``\textit{[...] There are experts in the company that help us understand and translate system architecture.}''

We also asked practitioners about their strategies to decompose their legacy systems into microservices. The results indicate that the most prevalent approach is \textit{decomposing the system by business capabilities, defining the responsibility of each microservice}, chosen by 28 (52.8\%) participants. This approach aligns with report studies on monolith refactoring~\cite{balalaie2016}, which advocate for organizing systems by business capabilities. Some papers suggest that each microservice should have one and only one business responsibility \cite{velepucha2023}.
An \textit{incremental strategy, focusing on one microservice at a time}, was selected by 27 (50.9\%) participants. This method often involves applying patterns like the strangler fig pattern \cite{fritzsch2019}, allowing for the gradual replacement of monolithic components with microservices.   
Less commonly, some companies decomposed their systems \textit{based on system operations and their data} (20.8\%), \textit{using UML diagrams} (3.8\%), or \textit{specialized tools like service cutters} (1.9\%) to aid in decomposition.

Participants were then asked about the criteria used to decide which microservices to extract. The results showed \textit{scalability} was considered the most relevant criterion, with 40 (75.5\%) participants noting its importance. This is likely because enhancing scalability allows for the expansion of microservices without requiring modifications to other system parts \cite{meloca2018}. As Newman discusses \cite{newman2019monolith}, breaking processing into individual microservices enables independent scaling, allowing companies to cost-effectively scale only those parts that constrain load while scaling down or turning off less utilized microservices when not needed.
\textit{Requirements} (52.8\%) and other criteria related to modularity such as \textit{reuse} (43.4\%), \textit{cohesion} (37.7\%), and \textit{coupling} (35.8\%), were all notably mentioned. This indicates an interest in maintaining well-defined and loosely coupled microservices. These findings are supported by various studies \cite{carvalho2019, bucchiarone2018}. The possibility of lower coupling and higher cohesion on microservice-based systems is also noted several times in literature \cite{jamshidi2013, alshuqayran2016, bucchiarone2018}. 

% Less mentioned criteria included database tables (22.6\%), variability (11.3\%), and network overhead (9.4\%), composing more specific technical considerations in the extraction process.

For the analysis regarding the process of migrating their systems to microservices, among the 53 participants, 27 (50.9\%) opted to gradually migrate a system initially implemented in a different architecture to the microservice architecture. Conversely, 5 (9.4\%) participants decided to migrate immediately, while 18 (34\%) participants built their new systems using the microservice architecture from scratch.
The prevalence of gradual migration is consistent with studies such as the work by Fan \& Ma~\cite{fan2017}, in which the authors justify gradual migration to maintain system stability and solve problems as they come up and are identified during the migration process. For the companies who chose microservices from scratch, this can be attributed to a need for faster development cycles~\cite{newman2015building, Kirby2021}. Although developing a system using microservice architecture can have higher initial costs, the potentially poor quality of the legacy code makes understanding and reusing the implementation a laborious activity~\cite{Mishra2022}. Thus, practitioners mainly rely on the team's knowledge of the legacy systems instead of reusing legacy code.

% \textcolor{red}{For the companies who chose microservices from scratch, this can be attributed to the necessity for faster development cycles \cite{newman2015building, Kirby2021}. Although developing systems using microservice architecture can have higher initial costs, reusing microservices makes developing new applications much faster in the long run\cite{Mishra2022}.}

To investigate the activities involved in migrating from monolithic architectures to microservices, we asked participants about the specific steps taken during their migration process.
The most frequently cited steps were \textit{decomposing the legacy system and identifying microservices} (75.5\%) and \textit{understanding the legacy system} (71.7\%). This is supported by recommendations found in modernization literature \cite{Wolfart2021}, where understanding and decomposing the legacy system is foundational to the process \cite{taibi2017, velepucha2023}. These steps ensure that the migration is based first on comprehending the existing system and its functionalities.
\textit{Defining the microservice architecture base} (58.5\%), \textit{Monitoring microservices and infrastructure} (58.5\%), \textit{Assessing the criteria/need that will lead to migration} (54.7\%), and \textit{Verifying and validating microservices} (50.9\%) were all selected by over 50\% of companies. %These activities are also covered in the paper by Wolfart et al. \cite{Wolfart2021}.
 These activities are also well-documented in academic literature \cite{Wolfart2021, zhou2023} and indicate that, for the most part, companies are executing methodologies proposed by experts.

In an open-ended question, we asked participants about their concerns regarding microservices granularity. This question had 43 responses, which resulted in five major themes.
The first theme is \textbf{Defining Clear Responsibilities}. This emphasizes the importance of establishing clear boundaries for each microservice to avoid duplicated functionalities and ensure each service has a specific purpose. Participants mentioned the need for microservices to have well-defined responsibilities, such as P8’s focus on reducing coupling and ensuring clear service responsibilities: ``\textit{The main concern is defining the responsibility of each microservice and reducing coupling between them.}'' Participant P20 also said: ``\textit{Each microservice should perform one task and one task only.}'' This theme stresses the need for purposeful microservices that handle specific tasks, enhancing the overall efficiency and maintainability of the system.

The theme \textbf{Management/Integration and Communication Complexity} addresses the challenges of managing and integrating numerous microservices. Participants like P5 and P77 pointed out the difficulty in managing highly granular services, while P30 and P73 underscored the complexities in integrating these services, especially with multiple REST calls and communication issues. The increased complexity in managing a microservice ecosystem, as noted by P21 and P81, can lead to significant integration errors and communication problems, complicating the overall system architecture and maintenance. For instance, P82 noted: ``\textit{Problems with communication between the microservices, as more services exist, more complex is the communication between them.}''

The third theme, \textbf{Quality Attributes}, encompasses concerns about cohesion, reusability, and other quality attributes of software architecture. Participants mentioned the balance between reusability and performance, with P4 noting: ``\textit{The fundamental challenge is finding an adequate level to ensure reusability without compromising performance.}'' Network overhead and horizontal scalability were also important considerations, as discussed by P69 and P7, respectively. Concerns with Maintenance and load balancing can be seen in the responses of P38, ``\textit{Code maintenance/database structure, load balancing}'' and P74, ``\textit{High granularity can increase complexity, resulting in harder maintenance. }''. This shows the importance of ensuring the system remains efficient and manageable over time.

For the theme \textbf{Balancing Service Size}, we see the attention to avoiding extremes in microservice granularity. Participants such as P15 and P28 warned against creating ``nanoservices'' that are too small and difficult to manage, as well as avoiding distributed monoliths that are too large and complex. P15 said: ``\textit{We strive not to create nanoservices or distributed monoliths; we aim to create services focused on a business domain.}'' This same concept can be seen in this quote from participant P70: ``\textit{The concerns are to not implement too small microservices, which would lead to creating a huge number of services, and do not implement too large microservice, which will be considered a monolith.}''

The fifth theme, \textbf{Team and Resource Management}, encompasses the challenges of managing the human and resource aspects of microservice architecture. P25 highlighted the need for a large team to manage numerous microservices. Similarly, P11 discussed the need to keep microservice scopes manageable for small and multifunctional teams: ``\textit{At most, they [services] should have a scope that a cross-functional team of 5-8 can autonomously manage.}'' 
Effective team and resource management are critical for maintaining and scaling the microservice architecture appropriately without overwhelming development and operations teams.

For the last open question regarding the migration process, we asked participants whether their companies followed a defined process for defining new microservices. Our thematic analysis found four major themes in 42 responses.

The first theme concerns \textbf{Ad Hoc and Informal Processes}. Many participants reported a lack of a formal process for defining new microservices, relying instead on ad hoc and informal methods. For example, P73 said: ``\textit{It’s like an ad hoc process; we do whiteboard sessions to discuss the team ideas and define a structure for the new microservice if it does not fit in an existing one.}'' In total, 19 participants explicitly stated there was no formal process in place. P67 and P68 mentioned that while there was no formal process, engineers would propose system changes to management and periodically assess microservice quality.

Next, \textbf{Architecture Team’s Role} emphasizes the involvement of architecture teams in the microservice definition process. Participants like P2 noted that the architecture team analyzes requests from the product team to determine whether a new microservice is needed, or if an existing one can be enhanced: ``\textit{There is an architecture team that analyzes the request from the product team for a new functionality and evaluates whether it is necessary to create a new microservice or if this new function can/should be included in an existing one.}'' Yet, P43 mentioned that architecture evolution often leads to the extraction of microservices from existing monoliths.

The third theme, \textbf{Business Context and Needs}, connects microservice creation to business requirements. Participants shared that new microservices are often defined based on business contexts and needs. P38 mentioned that a well-defined business context is a strong indicator for implementing a new service. Similarly, P80 said: ``\textit{They are generated as the business needs.}'' Participant P81 gave three situations where his company usually creates new services: ``\textit{1) a new domain object is being introduced; 2) an existing service became too big, so it's being split out; 3) a new `action' (i.e., a set of functionalities) is being introduced in the system, and it doesn't logically fit into an existing service.}''

Lastly, we have \textbf{Other Methods} that includes various less-mentioned considerations, such as periodic quality assessments (P68) or monitoring production behavior and creating documentation (P72). Sprint review meetings (P3) and mapping and prioritizing manual processes (P16) were also mentioned as part of the microservice definition process.

\begin{opus-box}
\textbf{Answering RQ2:} 
Companies adopt a systematic and incremental approach to migrating monolithic legacy systems to microservices. The most common strategy used by 50.9\% of participants is \textit{decomposing systems by business capabilities}. %This makes sense as one of the main concerns mentioned by participants is defining clear service responsibilities. Key activities in the migration process include understanding the legacy system (71.7\%) and defining the microservice architecture (58.5\%). 
Scalability is the most critical criterion for defining microservice candidates, cited by 75.5\% of participants. Other significant criteria include requirements, reusability, cohesion, and coupling. Practical strategies for understanding legacy systems, such as involving experienced developers \cite{taibi2017} and domain-driven analysis \cite{fowler2020, velepucha2023}, are widely used and supported by the literature.
%Additionally, many companies have reported using ad hoc and informal processes to define new microservices.

\end{opus-box}

% -- -- -- -- -- -- -- -- -- -- -- -- -- -- -- -- --

\subsection{RQ3 - Database and Data Consistency Concerns}
\label{3.database}

Managing data while ensuring consistency is a recurrent challenge when migrating legacy systems to microservices, mainly because of the distributed nature of this type of architecture~\cite{newman2019monolith, wu2022}. Nonetheless, existing work on modernization with legacy systems falls short when the topic is data management. Thus, we asked participants several questions to understand how they handled database structures, transactions, and consistency in their microservice architectures.

% \subsubsection{Quantitative Analysis}

First, we inquired practitioners about the structure of the databases used in their microservice architectures. The majority, 31 (58\%), reported using a database dedicated to each microservice. Interestingly, this result slightly differs from a 2018 survey with practitioners~\cite{difrancesco2018}, in which more than half of participants mentioned using mainly one single database for all microservices.
Having one database per microservice works well with the principles of microservice autonomy, reducing coupling and enhancing scalability \cite{newman2019monolith}, as recommended by Lewis \& Fowler in their seminal article about microservices~\cite{Lewis2014}. However, decentralized data leads to the problem of guaranteeing the consistency of said data \cite{velepucha2021}.
Despite this recommendation, 10 (19\%) participants indicated using a centralized database accessed by all microservices. While convenient, this design decision can increase coupling and reduce the autonomy of microservices \cite{newman2019monolith}. Interestingly, while participants confirm using centralized databases in microservices to address data consistency issues~\cite{baskarada2020}, these participants acknowledge that a certain level of centralization is counterproductive to the spirit of the architecture~\cite{baskarada2020,newman2015building}.
\textit{Connection pooling}, a technique to improve database connection efficiency \cite{sobri2022}, was used by 40 (75\%) participants, reflecting the interest in maintaining the performance of microservice-based applications.

When asked whether their choice of microservices was influenced by the transactions they execute in the database, 25 (47\%) participants indicated that transactions were evaluated and refactored when necessary to allow for separation into microservices. Conversely, 10 (19\%) did not evaluate transactions, and 6 (11\%) evaluated but did not refactor them.

Regarding consistency strategies, 22 (42\%) practitioners mentioned they adopted application-controlled eventual consistency, while 16 (30\%) used atomic transactions in the database. The choice between these approaches reflects trade-offs between consistency and performance \cite{faustino2024}. A few companies, 5 (9.4\%), use both approaches case-to-case.

Handling database tables across multiple microservices is another point of interest, as it can lead to tightly coupled services. In our survey, 18 (34\%) participants reported that their microservice-based systems involve multiple services managing the same tables. According to Ntentos et al.~\cite{ntentos2019}, it is crucial to determine the data to be shared between microservices, the frequency of updates, and the primary data provider. These decisions ensure that data sharing does not result in tightly coupled services.
Interestingly, 36 (68\%) participants stated that their microservices do not share persistent business classes with each other, which supports the autonomy and decoupling principles of microservice architectures \cite{newman2015building}.

To the participants who said their microservices \textit{do} share persistent business classes, we asked an open question about what strategy was adopted to share classes of persistent entities used by more than one microservice. Our analysis of those responses resulted in two major themes.

The first theme, \textbf{Common Code Packages and Libraries}, addresses using shared packages and libraries to manage persistent entities across microservices. Participants like P9 and P12 mentioned using a ``common package'' and a shared Maven library, respectively, while others, such as P13, reported having a shared base project containing all entities: ``\textit{We have a base project shared between the microservices, this project contains all entities, even those used by only one microservice.}'' This theme highlights the importance of maintaining consistency and reducing redundancy in data handling across services. The theme of \textbf{Event Sourcing} comprises the utilization of event sourcing, as cited by P25, as a strategy to manage shared persistent entities.

The performance and scalability of databases were considered important by 31 (58\%) participants during the design of their microservice architectures. These findings reveal that companies actively address data persistence and consistency challenges while migrating to microservices. Although guidelines and practices are seen in the literature \cite{newman2019monolith, richardson2016-2}, persistence strategies vary. Issues like data consistency \cite{velepucha2021, baskarada2020} and the complexity of managing multiple databases \cite{kalske2018} still pose challenges.

% Here, we also asked two open questions. The first one was directed to the participants who considered the importance of the performance and scalability of their databases.

Our questionnaire included an open-ended question directed to participants who considered the performance and scalability of their databases important. We asked those participants about the strategies adopted to scale the database in the face of performance bottlenecks. Our analysis of those responses resulted in five themes.

The first theme is \textbf{Asynchronous Processing and Event Handling}, which includes strategies like CQRS (Command Query Responsibility Segregation), mentioned by participants P25 and P39, and asynchronous processing, noted by Participant P5. These methods help manage load and improve performance by separating read and write operations \cite{fowler2011}. The second theme, \textbf{Testing and Monitoring}, includes load testing and continuous monitoring, as highlighted by P10, P28, and P72. These practices ensure the system's scalability and performance are regularly assessed and optimized. The third theme is \textbf{Scalable Database Services}, which refers to using database services that inherently provide scalability, as pointed out by P15: ``\textit{Using DB services that already provide scalability.}''. The quote by P28, in special, conveys both the first and second themes well, offering an interesting insight into their company's approach: ``\textit{We opted to use a Database as a Service (DBaaS) with constant monitoring, from time to time we perform load tests in the systems to evaluate possible code or infrastructural adjustments.}''

The theme \textbf{Auto-scaling and Vertical/Horizontal Scaling}, covers strategies like auto-scaling (P19, P67, and P78), vertical scaling (P43 and P69), and horizontal scaling (P66 and P69). These methods allow the system to adjust resources dynamically based on demand. P69 mentioned a case-to-case approach for different types of customers: ``\textit{Although rare, bottlenecks can occur if the same customer (tenant) heavily uses different microservices that access the same DB. In the case of a recurrent tenant with such a use case, horizontally scaling the DB (Clusters, Data Replication, Sharing) might be fruitful in the long term. Otherwise, short-term, more simple vertical scaling of the DB servers.}''

Lastly, our fifth theme is \textbf{Code Optimization}, which refers to the need for code optimization to improve performance, as mentioned by the participants P68 and P70.

In the final open-ended question regarding database concerns, we asked participants who use centralized databases in their microservice-based systems what factors hindered their decision to partition the database to isolate the persistence of each microservice. This question had 19 responses, resulting in five major themes.

\textbf{Early Stage Migration Priorities} indicates that maintaining a centralized database is often a temporary measure during migration. Participants like P27 noted that they were in the early stages of migration and still using what they called a ``distributed monolith.'' P74 stated: ``\textit{Database will be partitioned eventually. It was [not partitioned] yet due to higher priority aspects of the architecture.}''

The second database concerns theme, \textbf{Proprietary and Legacy Dependencies}, encompasses challenges related to proprietary databases and legacy system dependencies. P5 and P8 mentioned being constrained by these factors. P8 reported: ``\textit{The database is shared by the legacy, and in particular by our BI [business intelligence].}''

We also observe financial constraints as considerable barriers in quotes leading to the third theme \textbf{Cost and Licensing Issues}. Participants like P12 and P19 mentioned the high licensing costs associated with new database instances.

In the fourth theme, \textbf{Context Separation and Complexity}, the feedback from P39 and P43 provides some interesting information covering the complexities of separating contexts within a monolithic database. P39 shared their approach to handling access to a centralized database: ``\textit{[...] access to a single database generally leads developers to choose the easier path, as the desired data can be `just a join away.' What I recommend for projects with good prospects for exponential growth is to segregate access into different users and schemas, minimally separating the domains and responsibilities of the application. This adds some complexity, but nothing compared to having another database, and it helps you keep responsibilities segregated.}''

The last theme found in our analysis is \textbf{Data Compliance, Consistency, and Integrity}. P43's answer is noteworthy as it addresses concerns of context separation and data compliance: ``\textit{The issue was the design of these microservices without considering what we now regard as best practices. Over time, the applications become complex and consequently difficult to segregate. Another challenge is maintaining consistency between these services.}'' P69 and P77 also mentioned the importance of maintaining and ensuring compliance when dealing with sensitive data.

\begin{opus-box}
\textbf{Answering RQ3:}
The main concerns regarding data persistence focus on database structure, consistency, and performance. Most participants (58\%) use a dedicated database for each microservice. %Conversely, 19\% use a centralized database.
%To address performance, 75\% of participants use connection pooling, and nearly half (47\%) evaluate and refactor database transactions for microservice compatibility. 
Consistency strategies vary, with 42\% adopting eventual consistency controlled by the application and 30\% using atomic transactions, highlighting trade-offs between consistency and performance. Regarding data sharing, 34\% of participants report multiple microservices managing the same table. While 68\% avoid sharing persistent business classes. %For those who share persistent entities, strategies include using common code packages and libraries and event sourcing.
Practical strategies for scaling databases include asynchronous processing, event handling, load testing, continuous monitoring, scalable database services, and auto-scaling. Cost, legacy dependencies, and complexity are major barriers to database partitioning.
\end{opus-box}

% -- -- -- -- -- -- -- -- -- -- -- -- -- -- -- -- --

\subsection{RQ4 - Challenges of Working with Microservices}
\label{4.challenges}

The migration from legacy systems to microservice architectures presents several challenges~\cite{vuckovic2020, kalske2018}. In this section, we explore the prevalent challenges encountered by companies during this modernization process.

% \subsubsection{Quantitative Analysis}

The first question to practitioners was about reactivating or reusing legacy code. Among the answers, 10 (19\%) participants stated that their companies had to reactivate or reuse some functionality from the legacy system. We then posed an open-ended question, asking participants to detail why they had to reactivate these functionalities. Our analysis of their responses identified three major themes.

The first theme is \textbf{Technical Issues in New Implementations}, such as bugs (P10), performance issues (P20), and integration problems (P26, P30, and P80). These issues often necessitate falling back to legacy systems. The second theme is \textbf{Strategy for Coexistence and Validation Periods}, for which P75 reported that their systems are designed to allow legacy and microservice architectures to coexist: ``\textit{What we do is design our system to allow both legacy and microservices to co-exist for a certain time. Once the solution is validated, the feature on the legacy system is shut down.}'' This approach is aligned with recommendations from \cite{newman2019monolith}.

For \textbf{Meeting Customer-Specific Needs}, participants like P26 and P83 mentioned that legacy functionality is sometimes retained to meet specific customer requirements during the transition. For instance, P83 stated: ``\textit{The legacy system was customized to use the new [microservices-based] one. This was done to prevent our client's operation from being interrupted [...] This part of legacy integration had a short lifespan, and new clients had to use the new system.}''

% Out of these 10 participants, 7 (70\%) explained that this measure was taken due to issues encountered while launching the migrated system into production. The remaining 3 (30\%) participants mentioned adopting a strategy where both systems (legacy and migrated) run simultaneously, processing requests in both and comparing their responses.}
% However, after a certain period of time, when the proper functioning of the migrated system could be ensured, the legacy system was deactivated, which aligns with recommendations from \cite{newman2019monolith}. Some of the issues mentioned for the reactivation include bugs in the new implementation, integration errors, design decisions, production environment performance issues, and third-party system integration problems.

Another challenge practitioners face when defining the microservice architecture is choosing how to approach reuse needs. In another question, we asked whether duplicate code exists in different microservices. For the 15 (28\%) participants who said \textit{yes}. Then, we asked them their experiences on handling duplicated code.

Our analysis of this open-ended question revealed two main themes. The first involves using \textbf{Shared Libraries and Common Services}. P3 said: ``\textit{Building libraries for code reuse,}'' and P26 highlighted the reuse of existing services: ``\textit{If we identify that the code exists in another microservice, and it makes sense to call that microservice to avoid duplicating the code, reuse is implemented.}''. This approach maintains consistency and reduces redundancy.

The second identified theme is \textbf{Independent Handling and Lack of Traceability}. P16 explained: ``\textit{Every time a change is made, it has to be done in both places and delivered together.}'' And P66 shared that duplicate code is ``\textit{handled separately in each microservice.}'' These are useful insights into why reuse is important in microservices contexts. Such approaches can lead to maintenance challenges, as there is no traceability or centralized management of duplicate code. For instance, P11 stated: ``\textit{There is no process or documentation for traceability of duplicate code.}''

Another challenge observed is regarding testing, which is an important aspect of software development. Organizations dedicate a lot of time and effort~\cite{ghani2019microservice}. Working with microservice architecture adds another level of complexity to that activity~\cite{richardson2016}. 
Therefore, participants were questioned about the types of tests performed during the development/migration of microservices in their companies. Our survey revealed that \textit{unit testing} (44 answers, 83\%) and \textit{integration testing} (42 answers, 79\%) are the most commonly used testing methods.
It makes sense that unit tests are preferred since they isolate the smallest usable components of an application before integration. This is particularly effective in microservices because all components are already separated~\cite{ghani2019microservice}. Also, according to Newman \cite{newman2015building}, microservices architecture follows standard software engineering practices for tests that happen before deployment.

Integration tests focus on identifying defects in the interactions and interfaces between components \cite{ghani2019microservice}. These tests ensure that each component behaves correctly when contacted and that API interfaces deliver the correct information in the appropriate format \cite{ghani2019microservice}. This level of testing, however, comes with its own set of microservice-specific challenges. Extensive system tests are impractical due to the higher frequency of releases \cite{heinrich2017}. Then, companies often rely on fine-grained monitoring techniques in production environments, allowing for the identification and quick correction of failures through new releases \cite{heinrich2017}. This fits the microservice principle of independently (re)deployable units \cite{newman2015building}.

Robust monitoring becomes essential to identify potential problems and ensure the ongoing performance and reliability of microservices in production~\cite{waseem2021design}. Thus, we asked participants whether their companies use automatic mechanisms for monitoring systems and what principles of microservice monitoring were used.
The responses revealed that a substantial majority (44 answers, 83\%) of the practitioners' companies employ automatic mechanisms for monitoring their systems. This monitoring includes checking the performance and scaling of the most utilized microservices, the level of coupling between different microservices, and the use of infrastructure resources.
In their study, Mazlami et al. \cite{Mazlamietal2017} emphasized the importance of robust logging and monitoring tools for achieving a mature level of microservice development. They also noted that 90\% of practitioners believe logging and monitoring should be established early in the projects to ensure effective oversight and management of microservice environments.

For the companies that use automatic monitoring, we further investigated the specific principles they apply. The results showed a diverse range of strategies: 36 (68\%) \textit{monitor both containers and what runs inside them}, and 36 (68\%) \textit{monitor APIs}. Additionally, 27 (51\%) utilize \textit{orchestration systems} to manage and monitor their microservices, while 22 (42\%) \textit{prepare for elastic and multinational services}. Finally, 12 (23\%) \textit{map monitoring to the organizational structure}.

We also questioned participants about the programming languages they used in their microservice-based systems. 25 (47.2\%) out of the 53 participants affirmed they used more than one programming language in their projects. This introduces a particular challenge: the need for instrumentation to monitor distributed systems across diverse technology stacks involving various programming languages and paradigms~\cite{heinrich2017}.
Furthermore, due to the frequent updates and changes inherent to microservices, defining a ``normal'' behavior becomes complex, complicating anomaly detection and often leading to false alarms \cite{heinrich2017}. These challenges highlight the need for robust monitoring strategies to maintain the performance and reliability of microservice environments.

Lastly, the final open question asked participants how their companies manage network overhead. This is an important topic since, unlike monolithic architectures, in microservices-based systems, communication between microservices can introduce significant performance overheads \cite{kalske2018}.

Our qualitative analysis revealed three main themes regarding challenges. The first theme is \textbf{Technical Solutions}, including caching (P10 and P12), load balancing (P10 and P43), and horizontal scaling (P43). Many participants (P15, P21, P25, P27, P39, P72, P74, P78, P79, P81, and P83) rely on cloud infrastructure to manage network overhead, as many current cloud services already offer a solution for this issue. P81 said: ``\textit{It mostly comes `for free' with AWS.}''

In the theme \textbf{Monitoring and Design Considerations}, the practitioners P5, P25, and P47 emphasized the importance of monitoring tools, while participants P11 and P67 mentioned changing protocol design and considering network overhead during microservice design.

An interesting theme that emerged in our analysis is that network overhead is \textbf{Not a Significant Concern}. 
Despite existing literature addressing network overhead during the migration~\cite{ICSME_tomicroservices, Carvalho2024}, for some participants (P4, P8, P13, P28, P66, and P71) the network overhead in microservices' communication is not an issue. Only a few participants noted it as a small concern. We see that in quotes by P20, who explained: ``\textit{We just recently started to monitor it [overhead], but it is not a big concern.}'' and P25 said that network overhead ``\textit{Is not a problem yet, but it is a resource being monitored.}''

\begin{opus-box}

% Few participants (19\%) reported reverting to legacy functionalities. Some mentioned bugs and integration problems. Usually, participants allow the legacy and microservices to coexist to validate their new system before shutting the old one down.

%30\% of the participants deal with duplicated code. To address that, most participants use shared libraries and reuse existing services.
\textbf{Answering RQ4:} Few participants (19\%) reported reverting to legacy functionalities, and 30\% of them dealt with duplicated code. To address this, most participants use shared libraries and reuse existing services. Unit testing (83\%) and integration testing (79\%) are the most commonly employed for testing. %However, extensive system tests are impractical due to frequent releases, making robust monitoring essential. 
Most of the companies (83\%) use automatic monitoring mechanisms. Common strategies include monitoring containers and APIs, using orchestration systems, and mapping monitoring to organizational structures. Network overhead is not a significant concern, with companies leveraging modern cloud infrastructure to ensure efficient communication and scalability across their microservices.

%To summarize, the most effective strategies to tackle the studied challenges include using shared libraries, designing for coexistence, leveraging cloud infrastructure, and employing robust monitoring tools.
\end{opus-box}

% ------------------------------------------------------
% ------------------------------------------------------
% ------------------------------------------------------

\section{Lessons for Practitioners}
\label{sec:lessons}

With our research questions answered, we can extract several practical lessons that can be useful for practitioners, which are presented in this section.

\textbf{Get your priorities straight:} prioritization when migrating is extremely important. Focusing on services with high maintenance or scalability issues can lead to significant early successes. This approach allows organizations to quickly benefit from the improved scalability and faster deployment cycles provided by a microservices architecture. Additionally, starting with less critical services can help teams refine their microservices strategy before dealing with more complex and vital parts of the system.

\textbf{Knowledge is power:} Understanding the existing system is paramount for effective modernization. Leveraging the knowledge of experienced developers and thorough documentation enables a more informed and controlled decomposition of the monolith. 

\textbf{Take small, calculated steps:} Incremental decomposition based on business capabilities allows organizations to manage the migration in smaller, less risky steps. Additionally, teams should focus on maintaining a balance between service granularity and complexity; overly granular or overly large services can create management and integration challenges that could negate the advantages of microservices.

\textbf{No silver bullets for managing data:} Data management in microservices presents its own set of challenges. While adopting a dedicated database per service is ideal for decoupling, it introduces complexity in maintaining data consistency. Strategies such as eventual consistency and atomic transactions provide different benefits depending on the system's requirements. Organizations should prioritize asynchronous processing, event-driven architectures, and continuous monitoring to manage scaling effectively. This approach, however, requires careful consideration of legacy dependencies, cost constraints, and the difficulties of database partitioning.

\textbf{Address code duplication proactively:} Practitioners should be aware of the challenges of managing duplicated code, which can arise during modernization. Companies often address this by utilizing shared libraries and reusing existing services, which helps maintain consistency and reduce redundancy across the system.

\textbf{Testing is not an afterthought:} The complexities of testing in a microservices environment require a strategic approach. Our participants employ a mix of unit and integration testing to ensure the system's reliability. This helps to validate individual services and their interactions, ensuring that the system functions as intended.

\textbf{Monitor everything, automate everywhere:} Robust monitoring, primarily through automated mechanisms, is essential for maintaining the health of microservices. Effective strategies include monitoring containers and APIs, using orchestration systems, and aligning monitoring practices with organizational structures to ensure comprehensive oversight.

\textbf{Leverage the cloud for network challenges:} While network overhead might seem a theoretical concern, most organizations find it manageable with modern cloud infrastructure. Cloud services offer built-in solutions for scalability and performance, highlighting the importance of a proactive monitoring approach and strategic use of cloud resources to mitigate performance challenges in microservice environments.

\section{Threats to Validity}
\label{sec:threats}

In this section, we present the threats to the validity of our work and their mitigation~\cite{Wohlin2000}.

\textbf{Internal Validity:} One potential threat is the diversity of participants and companies. Although our sample includes responses from various international practitioners, it may not represent a regular distribution across all regions. This diversity is needed for our study but also introduces variability in practices and experiences that may affect the results. However, our sample covers a range of companies with different sizes and domains, with practitioners based in different countries. %While different methodologies or terminologies could influence the findings, we believe the core experiences and challenges reported by the participants would remain consistent.

\textbf{External Validity:} Although diverse, we cannot claim that our survey sample represents all types of companies, considering factors such as size, domain, and geographical location%. This variability can impact 
, impacting the generalization of our results. To mitigate this threat, we distributed the questionnaire through various media and social networks to reach a broad audience.%, aiming to include participants from different backgrounds and company profiles.

\textbf{Construct Validity:} Several survey questions were close-ended with predefined responses, which might have led participants to choose these options instead of describing their specific situations. To address this, we validated them in a pilot study with experienced developers. The use of a higher number of close-ended questions aimed to reduce the complexity and time required to complete the survey, thereby minimizing participant fatigue. %Additionally, while interpreting the open-ended responses, we attempted to remain objective by using a thematic synthesis approach, though some bias may have inadvertently influenced our analysis.

\textbf{Conclusion Validity:} The conclusions drawn from our data rely on the accurate and honest reporting of experiences by participants. Any misreporting or misunderstanding of survey questions could affect the validity of our conclusions. We designed our survey to be clear and concise, conducting a pilot study to refine questions. Nevertheless, the self-reported nature of the data poses an inherent limitation.

% ------------------------------------------------------
% ------------------------------------------------------
% ------------------------------------------------------

\section{Conclusion}
\label{sec:conclusion}

This paper explores the motivations, activities, and challenges associated with migrating monolithic legacy systems to microservices. Our research aimed to understand how professionals conduct their migrations and handle the associated challenges, providing insights from 53 software practitioners across diverse international contexts.

Our main findings reveal that the primary driving forces for adopting microservices are operational and technical, focusing on enhancing maintenance, scalability, and deployment processes. Unit and integration testing, alongside extensive monitoring, are essential for managing dynamic microservice environments. Database management continues to be an issue, with companies divided between centralized and decentralized approaches. %While decentralized databases are preferred for their autonomy and scalability benefits, ensuring data consistency across services remains challenging. 
Additionally, Modern cloud technologies are used to mitigate network overhead in microservice architectures.

As contributions, our study provides a comprehensive view of the current practices and challenges in migrating to microservices from the perspective of industry. The diverse approaches and ongoing challenges reveal the need for continued research and tailored solutions to support successful microservice adoption.

% ------------------------------------------------------
% ------------------------------------------------------
% ------------------------------------------------------

\section*{Acknowledgment}
This work is supported by CNPq grant \mbox{404027/2023-7}, %thelma
FAPERJ PDR-10 program 202073/2020, %wesley
and CAPES - Finance Code 001.

% \clearpage
\bibliographystyle{ieeetr}
\bibliography{references}

\end{document}